\documentclass[11pt]{article}
\usepackage{epsfig,amssymb,graphicx,amsmath,amsthm,subcaption}
\usepackage{tikz}
\usepackage{float}
\usetikzlibrary{shapes.geometric, arrows}
\usetikzlibrary{positioning}
\usepackage{enumerate}
\usepackage{calrsfs}

\newtheorem{theorem}{Theorem}[section]
\newtheorem{lemma}[theorem]{Lemma}

\usepackage[mathcal]{eucal}

\begin{document}

\title{Time-delayed SIS epidemic model with population awareness}

\author{G.O. Agaba,\hspace{0.5cm}Y.N. Kyrychko,\hspace{0.5cm}K.B. Blyuss\thanks{Corresponding author. Email: k.blyuss@sussex.ac.uk} 
\\\\ Department of Mathematics, University of Sussex, Falmer,\\
Brighton, BN1 9QH, United Kingdom}

\maketitle

\begin{abstract}

This paper analyses the dynamics of infectious disease with a concurrent spread of disease awareness. The model includes local awareness due to contacts with aware individuals, as well as global awareness due to reported cases of infection and awareness campaigns. We investigate the effects of time delay in response of unaware individuals to available information on the epidemic dynamics by establishing conditions for the Hopf bifurcation of the endemic steady state of the model. Analytical results are supported by numerical bifurcation analysis and simulations.

\end{abstract}

\section{Introduction}

Recent outbreaks of communicable infectious diseases, such as Ebola, SARS, avian and swine influenza have highlighted an important role played by accurate reporting of disease cases and the global awareness campaigns in containing the outbreaks of these diseases and prevention of their subsequent re-appearance. This is also extremely important in the context of sexually transmitted infections, where the education campaigns have allowed to significantly reduce the disease incidence. Understandably, the spread of awareness can play both a positive role, resulting in the containment or eradication of a disease, and a negative role, 
as evidence by the failure of an HPV campaign in Romania due to negative press coverage \cite{Penta14}, or the spread of plague in one of the states in India due to panic and anxiety \cite{R01}. A number of mathematical models have looked at the roles of different factors associated with the simultaneous spread of disease and awareness, using a mean-field approach (see \cite{Agaba17,Funk09,Kiss10}, and \cite{Greenhalgh15,Manfredi13} for recent reviews of some of the existing models) or network \cite{Funk10b,Funk09,Funk10,Gross08,Hatzopoulos,Juher15,Sahneh11,Wang13,Wu12} models that can often provide a more detailed information about contacts between individuals.

Within the framework of mean-field models, there are two main approaches for including the spread of information into epidemic models. One possibility is to incorporate the effects of information directly into the disease transmission rates, so that the disease awareness would result in a reduced transmission of the
disease. This is usually represented in the form of an exponential \cite{Cui07,Liu07,Tchuenche12} or saturated \cite{Cui08,Li09,Sun11,Tchuenche11} growth of the multiplication factor. Another option is to explicitly introduce an additional compartment representing a level of disease awareness, so that transitions between unaware and aware classes of individuals would depend upon this new variable \cite{Misra11,Misra15,Samanta13}. Previous work on modelling the effects of disease awareness on the spread of epidemics has highlighted a number of important dynamical features, such as the emergence and co-existence of multiple feasible steady states \cite{Cui07,Liu07}, as well as the occurrence of multiple disease outbreaks \cite{Liu07}. It has also provided a methodology for analysis and development of optimal strategies for disease containment and prevention \cite{Li09,Roy15,Tchuenche11,Wang16}.

One of the practically important and epidemiologically relevant issues is the existence of a non-negligible time delays associated with reporting of infected cases and human response to available information about the disease. A number of models have looked into the effects of these time delays on the disease dynamics. Zuo et al. \cite{Zuo15} introduced time delay in the equation for the ``media" variable $M$ to account for the delay in reporting cases of infections, while Misra et al. \cite{Misra11jbs} have also included some degree of global awareness.  Zhao et al. \cite{Zhao14} incorporated delayed reporting into the reduced disease transmission rate. Zuo and Liu \cite{Zuo14} focused on the analysis of time delay between reports of infection and changes in the behaviour. In all these models, the disease-free steady state is stable when a basic reproduction number $R_0$ that depends on the disease parameters only satisfies the condition $R_0<1$, and for $R_0>1$, the disease-free steady state is unstable regardless of the value of the time delay. Also, for $R_0>1$, each of these models has a feasible endemic steady state that is stable for time delay equal to zero, and in the models of Zuo et al. \cite{Zuo15}, Zhao et al. \cite{Zhao14} and Misra et al. \cite{Misra11jbs} it can undergo a Hopf bifurcation at a certain value of the time delay. In the model of Zuo and Liu \cite{Zuo14}, the endemic steady state is globally asymptotically stable independently of the time delay, provided it is biologically feasible. Greenhalgh et al. \cite{Greenhalgh15} have included both the delay in reporting of infected cases, and another delay representing the loss of disease awareness after a fixed period of time. They have shown that increasing the duration of awareness leads to a reduced equilibrium of infected individuals, and both time delays can lead to a destabilisation of the endemic equilibrium and an onset of oscillations.

In this paper we analyse the dynamics of a simultaneous spread of infectious disease and awareness. We consider an SIS-type epidemic model and divide the total population, which is assumed to be constant, into susceptibles unaware of the diseases, whose proportion is denoted by $S_n$, susceptibles aware of the disease, whose proportion is denoted by  $S_a$, and infected individuals aware of the disease by virtue of being infected, whose proportion is denoted by $I$. The model focuses on a directly-transmitted infection with a disease transmission rate $\beta$, which is modified by a factor $0 < \sigma_s \leq 1$ in aware susceptibles to describe the prevention measures, such as, reduction in contact, use of vaccine etc., that they undertake in the light of the disease awareness. Once infected, individuals recover at a rate $r$ and return to the class of susceptibles (the disease is assumed to confer no immunity), with a proportion $p$ of them being aware of the disease, and proportion $q$ remaining unaware, so  that $p+q=1$. Disease awareness is lost at a rate $\lambda$, so the effective duration of awareness is $1/\lambda$.

The level of awareness in the population is denoted by $M$, and it contains a contribution from some global sources, such as, general public awareness and media campaigns represented by the constant value of $\omega_o$, global awareness stemming from the number of reported cases of disease, which is proportional to $I$ with a rate $\alpha_o$, as well as an input from aware susceptible individuals, taken to be proportional to $S_a$ with a rate $\alpha$. Once awareness starts to spread, unaware susceptible individuals become aware at a rate $\eta$, and the awareness is lost at a rate $\lambda_o$. To account for the fact  that even in the presence of information, it takes some time for individuals to actually become aware and modify their behaviour, we explicitly include time delay $\tau$ from the moment information becomes available to the time susceptible individuals process it, change their behaviour accordingly, and can be considered fully aware susceptible individuals.

With these assumptions, the model has the form
\begin{equation}\label{eqn3}
\begin{array}{l}
\displaystyle{\dot{S}_n =  -\beta I S_n- \eta M(t - \tau) S_n + \lambda S_a + r q I,}\\\\
\displaystyle{\dot{S}_a =-\sigma_s \beta I S_a + \eta M(t - \tau) S_n - \lambda S_a + r p I,}\\\\
\displaystyle{\dot{I}=\beta I S_n + \sigma_s \beta I S_a - r I,}\\\\
\displaystyle{\dot{M} =\omega_o + \alpha_o I +\alpha S_a - \lambda_o M,}
\end{array}
\end{equation}
with the initial conditions
\begin{equation}\label{eqn4}
\begin{array}{l}
S_n(0) = S_{n_0} \geq 0, \quad S_a(0) = S_{a_0} \geq 0, \quad I(0) = I_0 > 0,\quad S_{n_0}+S_{a_0}+I_0=1,\\\\
M(s) = \phi(s) \geq 0,\quad -\tau\leq s<0,\quad M(0) = M_0 \geq 0,
\end{array}
\end{equation}
Since this model has no vital dynamics or disease-induced deaths, the total population is constant, and $S_n(t)+S_a(t)+I(t)=1$. Before proceeding with the analysis, we have to ascertain  that solutions of the model (\ref{eqn3}) remain biological feasible for all $t \in [0, \infty)$.

\begin{theorem}
\label{thm1}
The solutions, $ S_n(t), S_a(t), I(t), M(t)$, of the system of equations (\ref{eqn3}) with initial conditions (\ref{eqn4}) are non-negative for all $t \geq 0$.
\end{theorem}

\noindent This result can be proven using standard techniques, and it also follows from Theorem 5.2.1 in \cite{Smith95}. Thus, we conclude  that during their evolution, solutions of the system (\ref{eqn3}) with initial conditions (\ref{eqn4}) will remain within the bounded set
\begin{eqnarray*}
\Phi  = \left\{(S_n, \; S_a, \; I, \; M) \; \in \; \mathbb{R}_+^4: \;  0 \; \leq S_n, \; S_a, \; I \; \leq \; 1, \; 0\leq M\leq\widetilde{M}\right\},
\end{eqnarray*}
where
\[
\displaystyle{\widetilde{M}=\max\left[M_0,\frac{\omega_o+\alpha_o+\alpha}{\lambda_o}\right].}
\]

The outline of the paper is as follows. In the next section we establish conditions for feasibility of the steady states of model (\ref{eqn3}) and determine their stability. We identify conditions for Hopf bifurcation of the endemic steady state in terms of system parameters and the time delay. Section 3 contains results of numerical computation of characteristic eigenvalues, as well as numerical bifurcation analysis and direct numerical simulations. The paper concludes with discussion of results in Section 4.

\section{Steady states and their stability}

It is straightforward to show  that for any values of parameters, the system (\ref{eqn3}) has a {\it disease-free steady state} $E_0=(S_n^0, S_a^0, 0, M^0)$,  where
\begin{eqnarray}
\label{eqn6}
S_n^0=  1 - h_o, \qquad S_a^0 = h_o, \qquad  M^0= \frac{\omega_o + \alpha h_o}{\lambda_o},
\end{eqnarray}
with
\begin{equation}\label{hodef}
\displaystyle{h_o =  \frac{1}{2}\left(1 - \frac{\lambda \lambda_o + \eta \omega_o}{\eta \alpha} \right) + \sqrt{\frac{1}{4} \left(1 - \frac{\lambda \lambda_o + \eta \omega_o}{\eta \alpha} \right)^2 + \frac{\omega_o}{\alpha}}}.
\end{equation}
One should note that the since $0<h_o<1$ for any $\omega_o>0$, in this case the disease-free steady state is biologically feasible for any values of parameters, and in the absence of general awareness campaigns, i.e. for $\omega_0=0$, $E_0$ is only feasible, provided
\[
\eta\alpha>\lambda\lambda_o.
\]

\noindent The system (\ref{eqn3}) also has an {\it endemic equilibrium} $E^*=(S_n^*, S_a^*, I^*, M^*)$ with
\begin{equation}\label{eqn12}
\begin{array}{l}
\displaystyle{S_n^* =  \frac{x_2 \pm \sqrt{x_2^2 - 4 x_1 x_3}}{2 x_1}, \qquad I^* = \frac{r \lambda \lambda_o + \beta \eta \alpha S_n^{*2} -  (\beta \lambda \lambda_o + \sigma_s \beta \eta \omega_o + r \eta \alpha) S_n^*}{\sigma_s \beta [(\eta \alpha_o + \beta \lambda_o) S_n^* - r q \lambda_o]},}\\\\
\displaystyle{S_a^* = \frac{r - \beta S_n^*}{\sigma_s \beta}, \qquad M^* = \frac{\omega_o + \alpha_o I^*+\alpha S_a^*}{\lambda_o},}
\end{array}
\end{equation}
where
\[
\begin{array}{l}
x_1 = \beta [(1 - \sigma_s) (\eta \alpha_o + \beta \lambda_o) - \eta \alpha],\\\\
x_2 = \beta r q \lambda_o (1- \sigma_s) + (\eta \alpha_o + \beta \lambda_o) (r - \sigma_s \beta) - \beta (\lambda \lambda_o + \eta \sigma_s \omega_o) - r \eta \alpha,\\\\
x_3 = r \lambda_o [q (r - \sigma_s \beta) - \lambda],
\end{array}
\]
and for biological feasibility, the value of $S_n^*$ must lie within the interval
\begin{equation}\label{feas}
\frac{r q \lambda_o}{\eta\alpha_o + \beta \lambda_o} < S_n^* < \frac{r}{\beta}.
\end{equation}

Linearisation of the system (\ref{eqn3}) near any steady state $(\widehat{S}_n,\widehat{S}_a,\widehat{I},\widehat{M})$ has a characteristic matrix
\begin{equation}\label{Jac}
J =\begin{pmatrix}\vspace{2mm}
	- a_1 - a_6 & \lambda & - a_2 + r q & - a_0 a_4 \\ \vspace{2mm}
	 a_6 & - a_3 - \lambda & - a_5 + r p & a_0 a_4 \\ \vspace{2mm}
	 a_1 & a_3 & a_2 + a_5 - r &  0 \\ \vspace{2mm}
	0 & \alpha & \alpha_o & -\lambda_o 
\end{pmatrix},
\end{equation}
where
\begin{equation}\label{adef}
\begin{array}{l}
\displaystyle{a_0=e^{-k \tau}, \quad a_1=\beta \widehat{I}, \quad a_2=\beta \widehat{S}_n, \quad a_3=\sigma_s \beta \widehat{I}, \quad a_4=\eta \widehat{S}_n,}\\\\
\displaystyle{a_5=\sigma_s \beta \widehat{S}_a, \quad a_6=\eta \widehat{M},}
\end{array}
\end{equation}
and $k$ is the characteristic eigenvalue.

\begin{theorem}
\label{thm2}
The disease-free steady state $E_0$ of the system (\ref{eqn3}) is linearly asymptotically stable for all $ \tau \geq 0$ if $R_0<1$, unstable for $R_0>1$, and undergoes a steady-state bifurcation at $R_0=1$, where
\[
\displaystyle{R_0=\frac{\beta (1 + \sigma_s h_o - h_o)}{r}.}
\]
\end{theorem}

\noindent{\bf Proof.} Evaluating the characteristics polynomial at the disease-free steady state $E_0=(S_n^0, S_a^0, 0, M^0)$ yields the following equation for characteristic eigenvalues $k$
\[
k(k + r- a_2 - a_5)\left[k^2 + k(\lambda_o + \lambda + a_6) + \lambda_o(\lambda  + a_6) - \alpha a_0a_4\right] = 0.
\]
One of the eigenvalues is always $k_1=0$, another is given by
\[
k_2=a_2+a_5-r=\beta(S_n^0+\sigma_s S_a^0)-r,
\]
and the rest are determined by the roots of the transcendental equation
\begin{equation}\label{k_trans}
\displaystyle{k^2 + k (\lambda_o + \lambda + a_6) + \lambda_o (\lambda  + a_6)-\alpha a_4e^{-k \tau}=0.}
\end{equation}
The eigenvalue $k_2$ is negative, provided
\[
\beta(S_n^0+\sigma_s S_a^0)-r<0\quad\Longleftrightarrow\quad\frac{\beta(S_n^0+\sigma_s S_a^0)}{r}<1,
\]
which, using the values of $S_n^0$ and $S_a^0$ from (\ref{eqn6}), can be recast as
\[
R_0=\frac{\beta (1 + \sigma_s h_o - h_o)}{r} <1.
\]
It is clear that when $R_0$ passes the value of $1$, the eigenvalue $k_2$ goes through zero and becomes positive, thus making the steady state $E_0$ unstable by means of a steady-state bifurcation.

For $\tau = 0$, the equation (\ref{k_trans}) turns into a quadratic
\begin{equation}\label{k_quad}
k^2 + k (\lambda_o + \lambda + a_6) + \lambda_o (\lambda  + a_6) -  \alpha a_4 = 0,
\end{equation}
whose roots are both negative if and only if
\begin{equation}\label{lam_M0}
\displaystyle{\lambda_o (\lambda  + a_6) >  \alpha a_4 \qquad \Longleftrightarrow \qquad \lambda_o (\lambda  + \eta M^0) > \eta \alpha S_n^0.}
\end{equation}
Substituting the values of $M^0$ and $S_n^0$ from (\ref{eqn6}) shows that this condition is equivalent to
\[
\lambda \lambda_o + \eta \omega_o + 2 \eta \alpha h_o > \eta \alpha \qquad \Longleftrightarrow \qquad
h_o > \frac{1}{2}\left(1 - \frac{\lambda \lambda_o + \eta \omega_o}{\eta \alpha} \right),
\]
which always holds in the light of (\ref{hodef}). Hence, for $\tau=0$ both roots of the equation (\ref{k_quad}) always have negative real part.

To investigate whether the disease-free steady state can lose its stability for $\tau > 0$, we first note that $k=0$ is not a solution of this equation (this follows immediately from (\ref{lam_M0})), so we look for solutions of the equation (\ref{k_trans}) in the form
$k = i\mu$. Separating real and imaginary parts gives the following system of equations
\[
\begin{array}{l}
-\mu^2+\lambda_o (\lambda  + a_6)= \alpha a_4 \cos(\mu \tau),\\\\
\mu (\lambda_o + \lambda + a_6)= - \alpha a_4 \sin(\mu \tau).
\end{array}
\]
Squaring and adding these two equations yields a quartic equation for $\mu$
\[
\mu^4 + y_1 \mu^2 + y_2 = 0,
\]
with
\[
y_1 = \lambda_o^2 + (\lambda  + a_6)^2,\quad y_2 = (\lambda_o (\lambda  + a_6) + \alpha a_4) (\lambda_o (\lambda  + a_6) - \alpha a_4).
\]
Since $y_1 > 0$, and $\lambda_o (\lambda  + a_6) > \alpha a_4 $, which means that $y_2 > 0$, this suggests that there are no real positive roots $\mu^2$ of the above equation, such that $k = i \mu$ would be a root of equation (\ref{k_trans}). Consequently, the disease-free state is always stable if $R_0< 1$ for all $\tau \geq 0$.\hfill$\blacksquare$

\vspace{0.3cm}
Next, we investigate the stability of the endemic equilibrium of system (\ref{eqn3}). Evaluating the Jacobian (\ref{Jac}) at the endemic equilibrium $E^*=(S_n^*,S_a^*,I^*,M^*)$ gives the following characteristic equation for eigenvalues $k$:
\begin{equation}\label{char_end}
k \Big(k^3 + k^2 (\lambda_o + g_2) + k (\lambda_o g_2 + g_3 - \alpha a_0 a_4) + \lambda_o g_3 + a_0 [\alpha a_4(a_1 - a_3) -  \alpha a_4 (a_1 + g_1)] \Big) = 0,
\end{equation}
where
\[
\begin{array}{l}
g_1 = r - a_2 -  a_5,\\\\
g_2 = \lambda + a_1 + a_3 + a_6 + g_1,\\\\
g_3= a_3 (a_5 + a_6) +  r p (a_1 - a_3) + a_1 (\lambda + a_3 - a_5) + g_1 (\lambda + a_3 + a_6).
\end{array}
\]
Substituting the values of the state variables from (\ref{eqn12}) shows that $a_5 = r - a_2$, and, furthermore,
\begin{equation}\label{a1a3}
\begin{array}{l}
a_1 - a_3 =\beta I^*(1 - \sigma_s)\geq 0,\\\\
\displaystyle{a_3 (a_1+a_5 + a_6)-a_1a_5=\frac{\sigma_s\,\beta\,I^*}{\lambda_o} \left[\eta(\omega_o + \alpha_oI^*) + \eta \alpha S_a^* + \beta \lambda_o I^* + \sigma_s \beta \lambda_o S_a^* - \beta \lambda_o S_a^*\right]>0,}
\end{array}
\end{equation}
which implies
\begin{equation}\label{g_def}
\begin{array}{l}
g_1=0,\qquad g_2= \lambda + a_1 + a_3 + a_6>0,\\\\
g_3=[a_3 (a_1+a_5 + a_6)-a_1a_5] +  r p (a_1 - a_3) + a_1\lambda>0.
\end{array}
\end{equation}
Hence, the characteristic equation (\ref{char_end}) simplifies into
\[
k\Big(k^3 + k^2 (\lambda_o + g_2) + k (\lambda_o g_2 + g_3 - \alpha a_0 a_4) + \lambda_o g_3 - \alpha a_0 a_3 a_4\Big) = 0,
\]
with one eigenvalue being $k = 0$, and the rest of the spectrum being given by the roots of the transcendental equation
\begin{equation}\label{end_trans}
k^3 + k^2 (\lambda_o + g_2) + k (\lambda_o g_2 + g_3) + \lambda_o g_3 = \alpha a_4 (k + a_3)e^{-k \tau}.
\end{equation}
For $\tau = 0$, the equation (\ref{end_trans}) turns into a cubic equation
\begin{equation}\label{eqn18}
k^3 + k^2 (\lambda_o + g_2) + k (\lambda_o g_2 + g_3 - \alpha a_4) + \lambda_o g_3 - \alpha a_3 a_4= 0.
\end{equation}
By the Routh-Hurwitz criterion, the roots of this equation have negative real part if and only if the following conditions are satisfied
\begin{equation}\label{RHcon}
\begin{array}{l}
\lambda_o + g_2 > 0, \qquad \lambda_o g_2 + g_3 - \alpha a_4 > 0, \qquad \lambda_o g_3 - \alpha a_3 a_4 > 0,\\\\
(\lambda_o + g_2)(\lambda_o g_2 + g_3 - \alpha a_4)>\lambda_o g_3 - \alpha a_3 a_4.
\end{array}
\end{equation}
Since $\lambda_o + g_2 = \lambda_o +  \lambda + a_1 + a_3 + a_6 > 0$, the first of these conditions is always satisfied. Using (\ref{g_def}), one has
\[
\lambda_o g_2 + g_3 - \alpha a_4 >\lambda_o g_2- \alpha a_4 =\lambda_o (a_1 + a_3 + a_6)+(\lambda\lambda_o-\alpha a_4)>0,
\]
where, due to the feasibility condition (\ref{feas}), one has
\begin{equation}\label{lamlam0}
\lambda\lambda_o- \alpha a_4>0,
\end{equation}
so the second Routh-Hurwitz condition in (\ref{RHcon}) also holds.

The third condition has the explicit form
\begin{equation}\label{RH3}
\lambda_o g_3 - \alpha a_3 a_4=\lambda_o[a_3 (a_1+a_5 + a_6)-a_1a_5]+
(\lambda_o r p+\alpha a_4)(a_1-a_3)+a_1(\lambda\lambda_o- \alpha a_4)>0,
\end{equation}
where all brackets in the last expression are positive due to (\ref{a1a3}) and (\ref{lamlam0}). Hence, we have the following result.\\

\begin{lemma}
If the condition
\begin{equation}\label{RH_final}
(\lambda_o + g_2) (\lambda_o g_2 + g_3 - \alpha a_4) > \lambda_o g_3 - \alpha a_3 a_4,
\end{equation}
holds, the endemic steady state $E^*$ is linearly asymptotically stable for $\tau=0$.
\end{lemma}

\noindent {\bf Remark 1.} {\it Although it does not appear possible to analytically prove that the condition (\ref{RH_final}) always holds, numerical simulations suggest that it does indeed hold for any parameter values, for which the endemic steady state $E^*$ is biologically feasible}.\\

Since we have now established that for $\tau = 0$ the endemic state $E^*$ is linearly asymptotically stable, one still has to find out whether this steady state can lose its stability for $\tau > 0$. First of all, one should note that in the light of inequality (\ref{RH3}), $k=0$ is not a root of the characteristic equation (\ref{end_trans}). Hence, the only way how the steady state $E^*$ can lose its stability is when a pair of complex conjugate eigenvalues crosses the imaginary axis from left to right. Introducing auxiliary parameters,
\[
y_3 = \lambda_o + g_2,\quad y_4 = \lambda_o\,g_2 + g_3, \quad y_5 = \lambda_o g_3,\quad y_6 = \alpha a_4,\quad y_7 =  \alpha a_3a_4,
\]
the characteristic equation (\ref{end_trans}) can be recast in the form
\begin{equation}\label{eqn20}
k^3 + y_3 k^2 + y_4 k + y_5 = (y_6 k + y_7) e^{-k \tau}.
\end{equation}
Substituting $k = i \mu$ into this equation and separating real and imaginary parts gives
\begin{equation}\label{eqn21}
\begin{array}{l}
- y_3 \mu^2 + y_5= y_7\cos(\mu\tau)  + y_6\mu\sin(\mu\tau),\\\\
- \mu^3 + y_4\mu=  y_6\mu\cos(\mu\tau) - y_7\sin(\mu\tau).
\end{array}
\end{equation}
Squaring and adding these two equations yields the following equation for the Hopf frequency $\omega$:
\[
f(\mu)=\mu^6 + (y_3^2 - 2\,y_4)\mu^4 + (y_4^2 - y_6^2 - 2y_3y_5)\mu^2 + y_5^2 - y_7^2=0,
\]
Without loss of generality, let us assume this equation has six distinct positive real roots $\mu_i$, $i=1,\ldots,6$. For each $\omega_i$, solving the system (\ref{eqn21}) for $\tau$, we find
\[
\tau_{j,n}=\frac{1}{\mu_j}\left[\arccos\left( \frac{y_5y_7 + (y_4y_6 - y_3y_7)\mu_j^2 -  y_6\mu_j^4}{y_7^2 + y_6^2\mu_j^2} \right) + 2\pi (n-1)\right],\; j=1,\ldots,6,\; n\in\mathbb{N}.
\]
\begin{figure}[!tb]
	\hspace{-0.8cm}
	\includegraphics[width = 19cm]{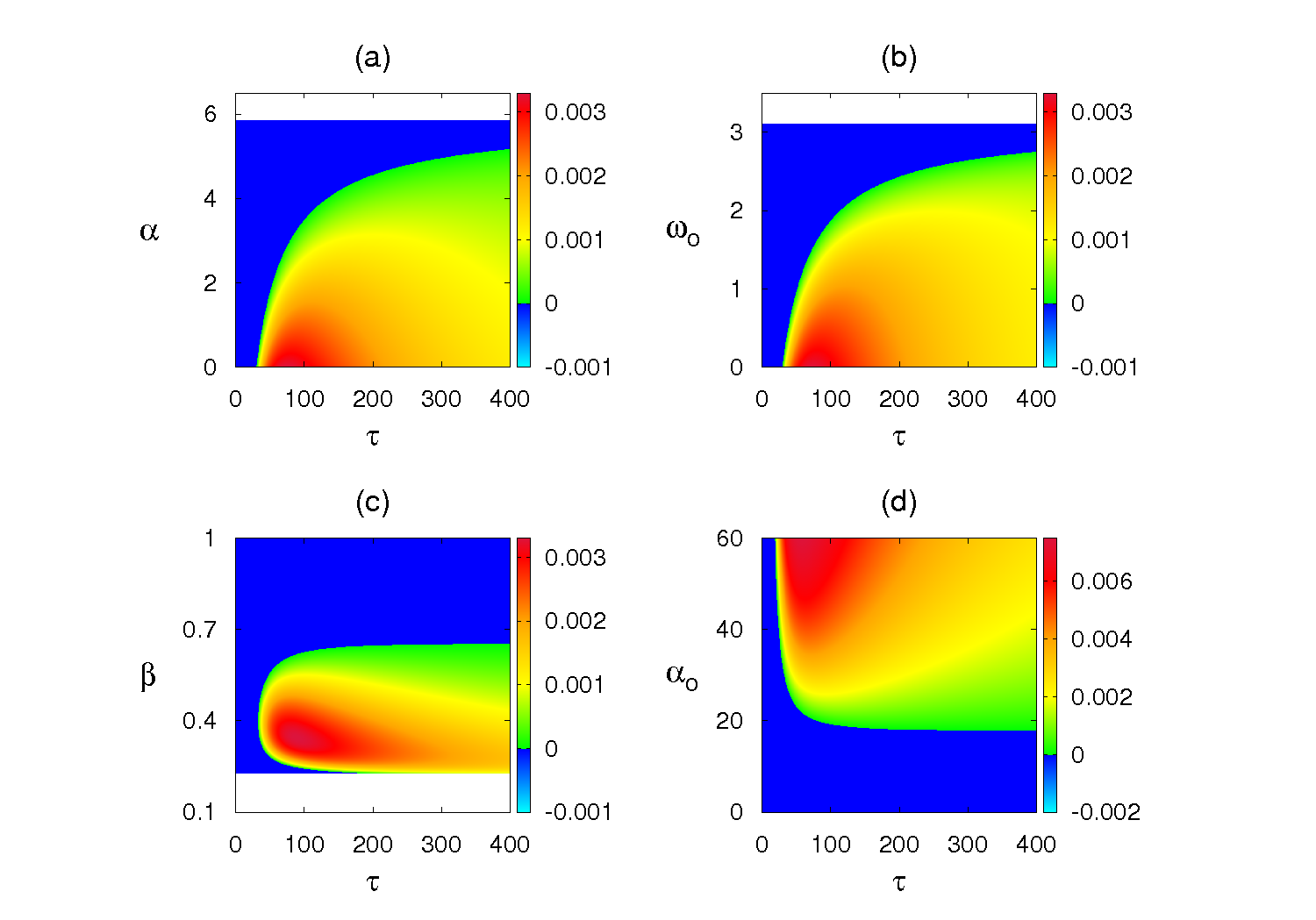}
	\vspace{-0.5cm}
	\caption{Stability of the endemic steady state $E^*$. Colour code denotes max[Re$(k)$] whenever the endemic steady state is feasible. Parameter values are as follows, (a) $\omega_o = 0.2, \beta = 0.4, \alpha_o = 30$; (b) $\alpha = 0.3, \beta = 0.4, \alpha_o = 30$; (c) $\alpha = 0.3, \omega_o = 0.2, \alpha_o = 30$; (d) $\alpha = 0.3, \omega_o = 0.2,  \beta = 0.4$. Other parameter values are $\lambda = 0.1,  r = 0.2, \sigma_s = 0.04, p = 0.9, q = 0.1, \lambda_o = 0.3, \eta = 0.01$.}\label{bif1}
\end{figure}
This allows us to define
\begin{equation}\label{tau_mu0}
\displaystyle{\tau_0=\tau_{j_0,n_0}=\min_{1\leq j\leq 6, n\geq 1}\{\tau_{j,n}\},\quad \mu_0=\mu_{j_0}.}
\end{equation}
 In order to establish whether the endemic steady state $E^*$ actually undergoes a Hopf bifurcation at $\tau=\tau_0$, one has to compute the sign of $d[{\rm Re}(k)]/d\tau$. Differentiating the characteristic equation (\ref{eqn20}) with respect to $\tau$ gives
\[
\left(\frac{d k}{d \tau} \right)^{-1}=\frac{y_6e^{- k\tau} - 3k^2 - 2y_3k - y_4}{(y_6k^2 + y_7k)e^{- k\tau}} -  \frac{\tau}{k}.
\]
\begin{figure}
\hspace{0.5cm}
	\includegraphics[width = 15cm]{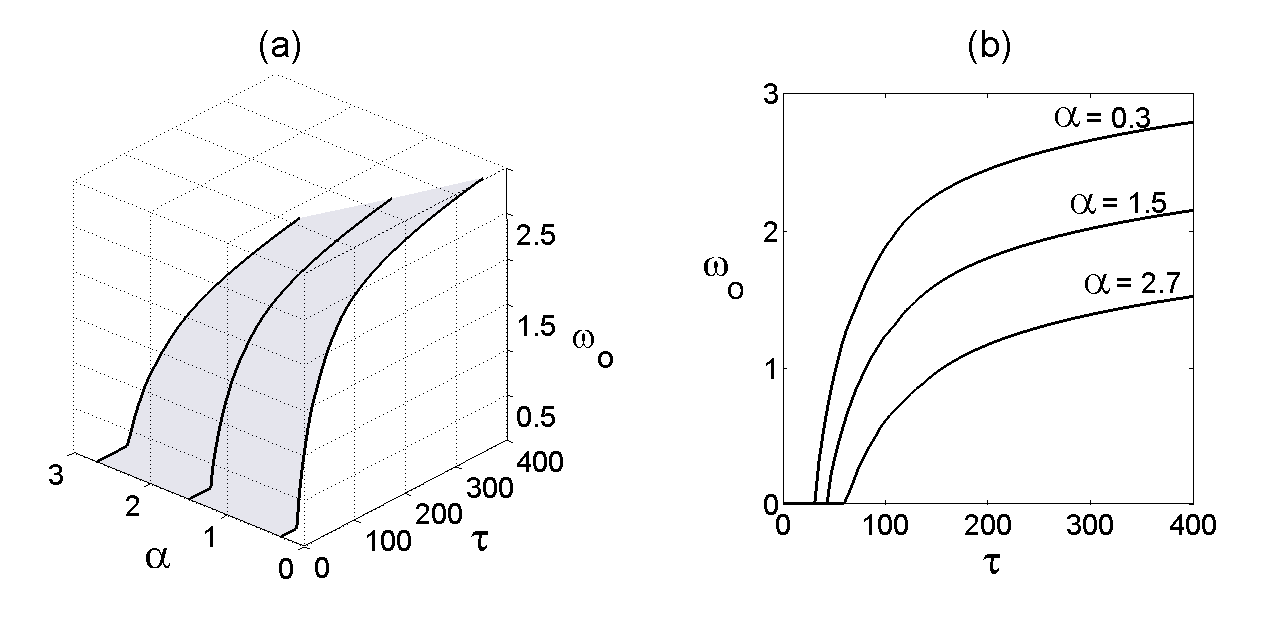}
\vspace{-0.6cm}
	\caption{Stability boundaries of the endemic steady state $E^*$. The steady state is stable to the left of the surface in (a), and to the left of the lines in (b).  Parameter values are $\lambda = 0.1, \beta = 0.4,  r = 0.2, \sigma_s = 0.04, p = 0.9, q = 0.1, \alpha_o = 30, \lambda_o = 0.3,  \eta = 0.01$.}\label{bif2}
\end{figure}
Evaluating this at $\tau=\tau_0$ with $k=i\mu_0$ and expressing $\sin(\mu_0\tau_0)$ and $\cos(\mu_0\tau_0)$ through coefficients $y_3,\ldots,y_7$ from (\ref{eqn21}) yields
\[
\text{Re} \left(\frac{d k}{d \tau} \right)^{-1} \Bigg|_{\tau=\tau_0} =\frac{3\mu_0^4 + 2(y_3^2 - 2y_4)\mu_0^2 - 2y_3y_5 + y_4^2 - y_6^2}{y_7^2 + y_6^2\mu_0^2} = z_0\,f^\prime(\mu_0),
\]
where $z_0=[2\mu_0(y_7^2 + y_6^2\mu_0^2) \big]^{-1}>0$. Hence, we have
\[
\text{sign} \left\{ \frac{d{\rm Re}[k(\tau_0)]}{d \tau} \right \}= \text{sign}\left\{ \text{Re} \left(\frac{d k(\tau_0)}{d \tau} \right)^{-1} \right \}=\text{sign}[k_0 f^\prime(\mu_0)]=\text{sign}[f^\prime(\mu_0)].
\]
These calculations can now be summarised in the following result.

\begin{theorem}
Let the conditions of {\bf Lemma 1} hold, and also let $\tau_0$ and $\mu_0$ be defined as in (\ref{tau_mu0}) and $f'(\mu_0)>0$. Then the endemic steady state $E^*$ is linearly asymptotically stable for $\tau<\tau_0$, unstable for $\tau>\tau_0$ and undergoes a Hopf bifurcation at $\tau=\tau_0$.
\end{theorem}

\section{Numerical stability analysis and simulations}

To get a better understanding of the effects of different parameters on the dynamics of the system, we expand the analysis presented in the previous section by numerically computing characteristic eigenvalues. This is achieved by using a pseudospectral method implemented in a traceDDE suite in MATLAB \cite{Breda06}. 

\begin{figure}
	\hspace{-0.7cm}
	\includegraphics[width = 18cm]{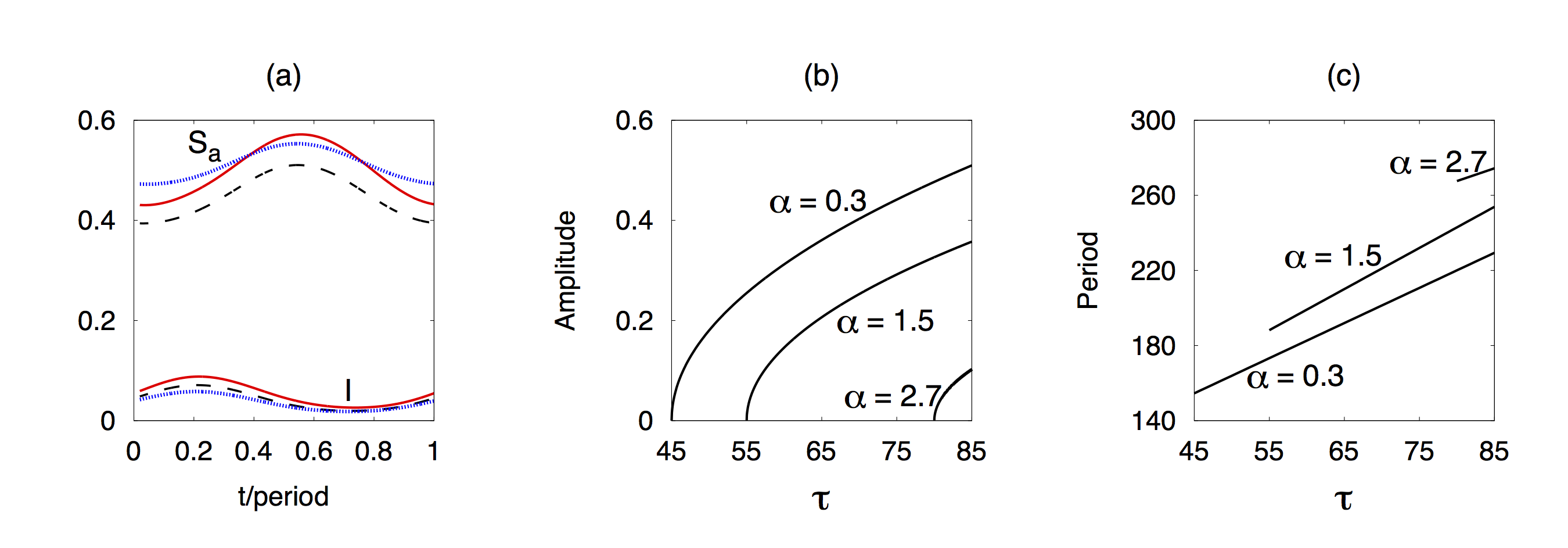}
	\vspace{-0.9cm}
	\caption{Bifurcation analysis of the endemic state: (a) periodic solutions: $\alpha = 0.3, \tau = 45$ (red solid lines), $\alpha = 1.5, \tau = 55$ (black dashed lines), $\alpha = 2.7, \tau = 80$ (blue dotted lines) and showing only the dynamics of $S_a$ and $I$; (b) plot of the amplitude against the time delay, $\tau$; (c) plot of the period against time delay, $\tau$. Parameter values are: $\lambda = 0.1, \beta = 0.4,  r = 0.2, \sigma_s = 0.04, p = 0.9, q = 0.1, \omega_o = 0.2, \alpha_o = 30, \lambda_o = 0.3, \eta = 0.01$.}\label{bif3}
\end{figure}

Figure~\ref{bif1} illustrates regions of stability of the endemic steady state $E^*$ depending on the disease transmission rate $\beta$, time delay $\tau$, and the rates of global, $\omega_0$ and local awareness $\alpha$, $\alpha_o$. This Figure shows that for sufficiently small time delays $\tau$, the endemic steady state $E^*$ is stable, thus providing numerical evidence to support {\bf Remark 1}. As the time delay $\tau$ increases, the steady state $E^*$ loses its stability in accordance with {\bf Theorem 3}. There are several important observations that have to be made here. First of all, one should note a qualitative difference in the effects of different types of awareness transmission. Whereas the endemic steady state $E^*$ can be destabilised for arbitrarily small values of the rates of the general awareness campaigns $\omega_o$, or local awareness $\alpha$, in the case of awareness associated with the increasing number of reported disease cases $\alpha_o$, the endemic steady state remains stable for all possible values of the time delay $\tau$.

Another counter-intuitive result is that as the rates of awareness $\omega_o$ and $\alpha$ increases, the endemic steady state actually remains stable for longer durations of the time delay in response. When one considers the effects of the speed of disease transmission, as shown in Fig.~\ref{bif1}(c), it becomes clear that for sufficiently high values of the disease transmission rate $\beta$, the endemic steady state is stable for any values of the time delay $\tau$. On the other hand, for small values of $\beta$, it has a destabilising role: as $\beta$ increases, the critical time delay at which the Hopf bifurcation occurs is decreasing, but this effect reverses starting with some value of $\beta$. For sufficiently large values of $\alpha$ and $\omega_o$, or for sufficiently small values of $\beta$, where $R_0<1$, the endemic steady state $E^*$ is not feasible, whereas the disease-free steady state $E_0$ is feasible and stable for any values of $\tau$.

Figure~\ref{bif2} shows how the stability boundary of the steady state $E^*$ changes depending on the rate of global awareness $\omega_o$ and local awareness $\alpha$. One can see that increasing the level of local awareness $\alpha$ results in the endemic steady state losing its stability for smaller values of the global awareness rate $\omega_o$ for the same time delay $\tau$. Conversely, if one fixes the value of $\omega_o$ and increases $\alpha$, instability occurs for higher values of $\tau$, suggesting that the local awareness actually helps the state of infection remain present in the population for longer durations of the individual response time.

\begin{figure}
\begin{center}
	\includegraphics[width = 9cm]{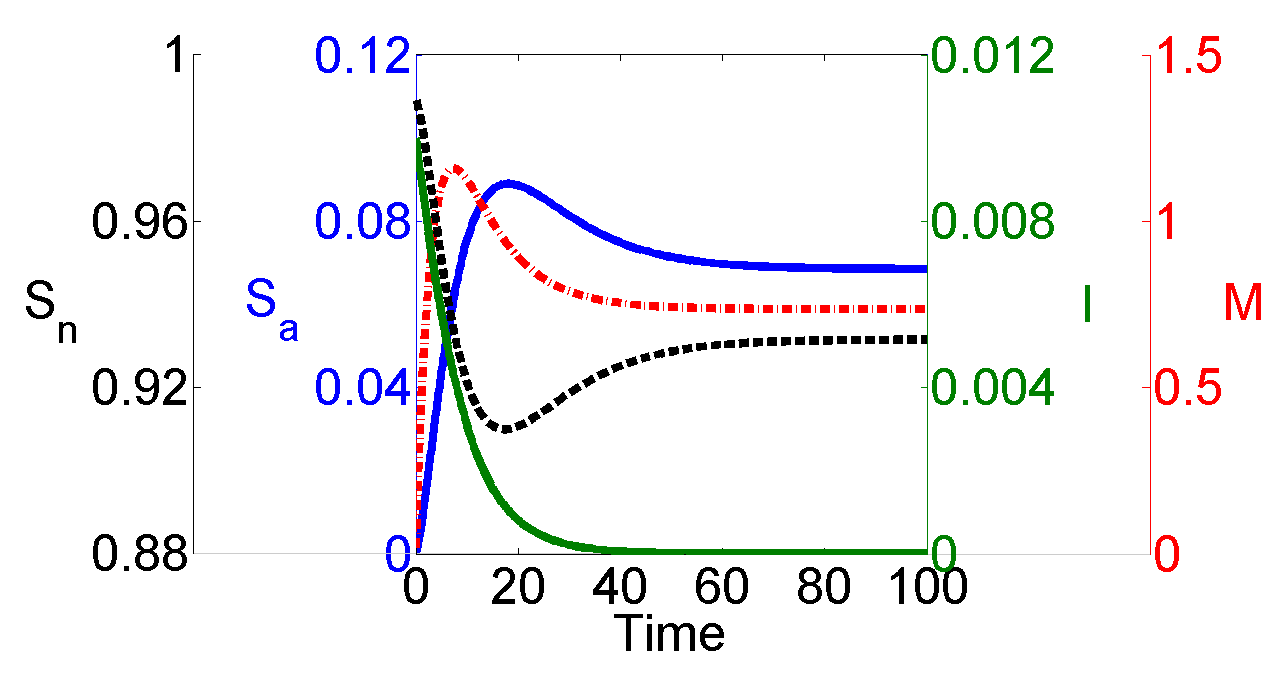}
	\caption{Numerical solution of the system (\ref{eqn3}) with $\tau=0$ and $\alpha = 0.3, \lambda = 0.1, r=0.5, \beta = 0.4, \sigma_s = 0.04, p = 0.9, q = 0.1, \omega_o = 0.2, \alpha_o = 30, \lambda_o = 0.3, \eta = 0.01$. In this case, $R_0=0.7474$, and the system settles on a stable disease-free steady state.}\label{df_sol}
\end{center}
\vspace{-0.5cm}
\end{figure}

To investigate the behaviour of the system beyond the Hopf bifurcation, we have used the continuation software DDE-BIFTOOL to numerically continue branches of periodic solutions in the parameter space, and the results are shown in Fig~\ref{bif3}. Figures~\ref{bif3}(b) and (c) indicate that increasing the time delay $\tau$ results in the larger amplitude and larger period of periodic oscillations around the endemic steady state $E^*$. For the same time delay $\tau$, the higher rate of local awareness $\alpha$ results in the smaller amplitude of oscillations, but a larger period of those oscillations, provided $\alpha$ is not too high to ensure the existence of periodic solutions.

\begin{figure}
	\includegraphics[width = 17cm]{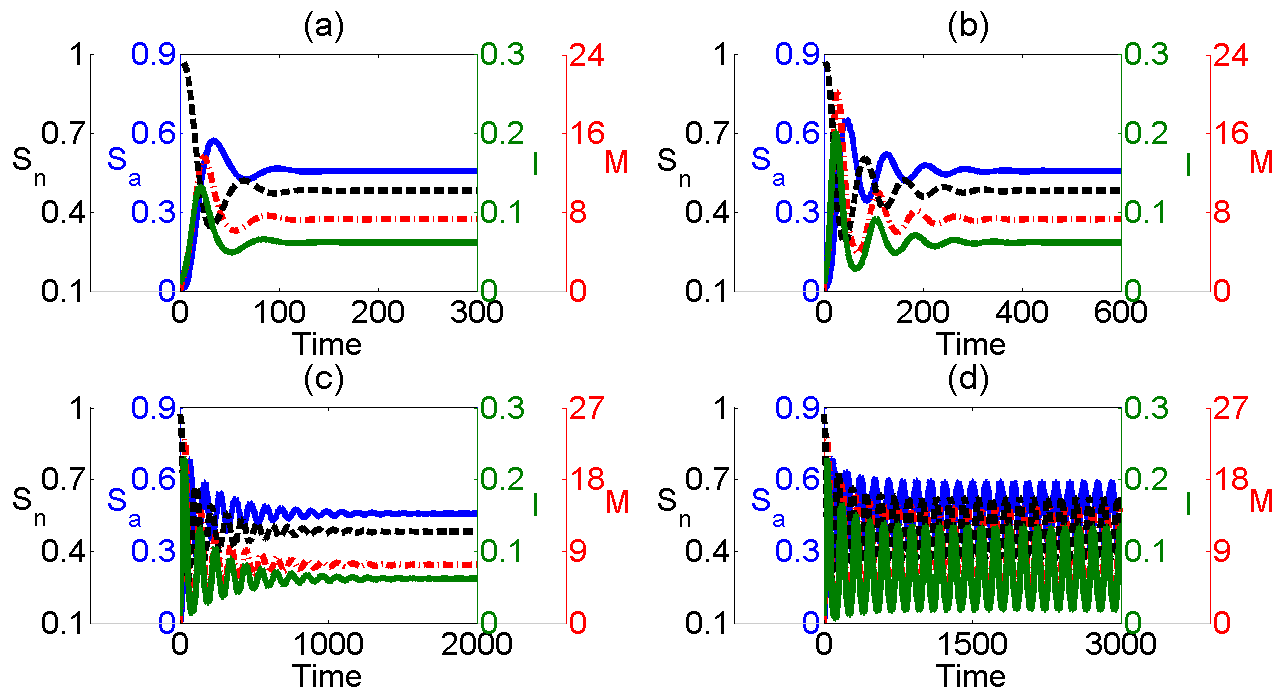}
	\caption{Numerical solution of the system (\ref{eqn3}) with $\alpha = 0.3, \lambda = 0.1, \beta = 0.4,  r = 0.2, \sigma_s = 0.04, p = 0.9, q = 0.1, \omega_o = 0.2, \alpha_o = 30, \lambda_o = 0.3, \eta = 0.01$ and $R_0=1.8685$. (a) $\tau = 5$, (b) $\tau = 14$, (c) $\tau = 25$, (d) $\tau = 42$.}\label{end_sol}
\end{figure}

Figure~\ref{df_sol} illustrates the dynamics of the system (\ref{eqn3}) in the case when $R_0<1$. In this situation, recovery from infection is sufficiently fast to ensure the initial outbreak is contained, and the disease is eradicated from the population. For slower recovery rates and sufficiently small delays in response to awareness, the system settles on a stable endemic steady state, as shown in Fig.~\ref{end_sol}(a)-(c). Increasing the time delay $\tau$ results in higher-amplitude decaying oscillations around this steady state, and once $\tau$ exceeds the critical value determined by {\bf Theorem 3}, the endemic steady state loses its stability, which results in the emergence of stable periodic solutions shown in Fig.~\ref{end_sol}(d).

\section{Discussion}

In this paper we have analysed the dynamics of a non-lethal infectious disease with the simultaneous spread of awareness, and a delayed response of individuals to available information. Specific emphasis was made on explicitly incorporating different facets of disease awareness that can arise due to general public information campaigns. This can be associated with the public reports of observed cases of the disease, or the spread as a word-of-mouth from aware to unaware individuals. We have derived conditions for feasibility and stability of the disease-free and endemic equilibria in terms of system parameters and the duration of the time delay associated with changes in individuals' behaviour. These results suggest that stability of the disease-free equilibrium is independent of the time delay but depends on the rates at which awareness is required. An interesting result is that for sufficiently high rates of the spread of global information or information from aware population, it is possible to eradicate the disease, whereas an increase in awareness stemming from the higher number of reported disease cases does not result in disease eradication. Another important observation is that in the presence of a delay in response of individuals to available information, increasing the rates of global or local awareness actually results in stabilising the endemic equilibrium, i.e. maintaining the disease presence in the population, and only when these rates get quite high, the disease is eradicated.

Considering the effects of the time delay on the disease dynamics, we have discovered that it can destabilise the endemic steady state, thus causing periodic oscillations. Both the amplitude and the period of these oscillations increase with the time delay in the individuals' response. The period is also growing with the rate of local information transmission, whereas the amplitude of oscillations decreases, and the oscillations can be completely suppressed for sufficiently high rates of local information transmission. If the disease is transmitted quite quickly, i.e. the transmission rate is sufficiently high, then increasing the delay will not affect the stability of the endemic equilibrium, hence, the disease will always be present at some constant level in the population. In a narrow range of values of the time delay, increasing the disease transmission rate initially destabilises the endemic steady state, whilst further increase leads to stability being regained.

The work presented in this paper provides some practical insights into the development and assessment of possible information campaigns targeted at disease control and prevention by elucidating how different routes of transmission of awareness affect the progression of the disease in the population. One possibility to make these results more realistic would be to consider an equivalent model of the simultaneous spread of the disease and awareness on a contact network with some realistic node degree distribution. There are significant theoretical and computational challenges associated with a non-Markovian nature of such models due to the presence of time delays, however, some progress has been recently made on the analysis of such models in the absence of awareness \cite{Kiss15}. Another related issue concerns including realistic distributions for the disease infectious and recovery periods, which can be effective represented by including multiple disease stages \cite{Sherb15s,Sherb15}, or through distributed delays \cite{Blyuss10,Wearing05}. Furthermore, one can realistically expect different individuals to respond differently to the information available to them, which can also be represented in the form of a delay distribution generalising the case of the discrete time delay considered in this paper.

\section*{Acknowledgements} GOA acknowledges the support of the Benue State University through TETFund, Nigeria, and the School of Mathematical and Physical Sciences, University of Sussex.

\bibliography{Literature}
\bibliographystyle{plain}

\end{document}